\documentclass[aps,pra,twocolumn,showpacs,draft,floatfix]{revtex4}
\usepackage{amsbsy}
\usepackage{amsmath}
\usepackage{epsf}
\usepackage{dcolumn}
\begin{document}

\title{A continuum of Hamiltonian structures for the two-dimensional
isotropic harmonic oscillator}

\author{Juan M. Romero}
\email{sanpedro@nuclecu.unam.mx}
\affiliation{Instituto de Ciencias Nucleares, UNAM, Apartado Postal 
70-543, M\'exico DF, M\'exico}
\author{Adolfo Zamora}
\email{zamora@nuclecu.unam.mx}
\affiliation{Instituto de Ciencias Nucleares, UNAM, Apartado Postal 
70-543, M\'exico DF, M\'exico}

\date{\today}

\begin{abstract}
We show the existence of a continuum of Hamiltonian structures for 
the two-dimensional isotropic harmonic oscillator. In particular,
a continuum of Hamiltonian structures having noncommutative 
coordinates is presented. A study of the symmetries of 
these structures is performed and their physical plausibility
is discussed.
\end{abstract}


\maketitle

It is well known that for some equations of motion there
exist inequivalent Hamiltonian/Lagrangian structures.
This is the so-called {\it inverse problem in variational calculus}
\cite{Maria:gnus}. It is also well known that these 
alternative structures yield different quantum theories 
\cite{wigner:gnus,ok:gnus,shepley:gnus,lemos:gnus,u:gnus}. 
In this work we present a continuum of Hamiltonian structures
for the two-dimensional isotropic harmonic oscillator;
in particular, a continuum of Hamiltonian structures with 
noncommutative coordinates. We also perform a study of their 
symmetries.\\

To begin with, let us recall that the usual Hamiltonian structure 
for the two-dimensional isotropic harmonic oscillator is given
by the Hamiltonian
\begin{equation}
H_0 = \frac{1}{2m}(p_{x}^{2}+p_{y}^{2}) + 
\frac{m\omega^2}{2} (x^2 + y^2),
\label{eq:H2}
\end{equation}
with Poisson brackets
\begin{equation}
\{ x , p_x \}_0 = 1, \quad \{ y , p_y \}_0 = 1,
\quad \hbox{\rm zero otherwise.} \label{eq:0}
\end{equation}
The rotational invariance is an important symmetry of the isotropic
harmonic oscillator. This symmetry is compatible with the usual
Hamiltonian structure as $H_0$ and the Poisson brackets from 
Eq. (\ref{eq:0}) are both invariant under rotations
\begin{eqnarray}
\begin{array}{ll}
& x' = x\cos\theta-y\sin\theta, \label{eq:ro1}\\
& y' = x\sin\theta+y\cos\theta. \label{eq:ro2}
\end{array}
\end{eqnarray}
Notice that this symmetry has $r^{2}=x^{2}+y^{2}$ as an invariant. 
Another property physically observable is that in the limit $\omega\to 0$
the oscillator becomes a free particle. Clearly the usual Hamiltonian
formulation above is also compatible with this.\\

Now, let us consider the Hamiltonian structure given by the
Hamiltonian
\begin{eqnarray}
H_{\alpha}&=& \Big[ \frac{1}{2m}(p_{y}^{2}-p_{x}^{2}) 
+\frac{m\omega^2}{2} (y^2 -x^2)\Big]\sin\alpha\nonumber\\
&-&\omega\Big[x p_y- yp_x\Big]\cos\alpha,
\label{eq:Haaa}
\end{eqnarray}
with Poisson brackets
\begin{eqnarray}
\begin{array}{lllll}
&\{ x , y \}_{\alpha} & = \cos\alpha/m\omega, 
&\quad \{ x , p_x \}_{\alpha} &= -\sin\alpha,\\
&\{ y , p_y \}_{\alpha} & = \sin\alpha, 
&\quad \{ p_x , p_y \}_{\alpha} & = m\omega\cos\alpha,\label{eq:aaa}
\end{array}
\end{eqnarray}
and zero otherwise. From a straightforward calculation it can
be shown that for every $\alpha$ the equations of 
motion are
\begin{eqnarray}
\begin{array}{lllll}
&\dot x & = {p_x}/{m}=\{ x , H_{\alpha}\}_{\alpha},  
&\dot p_x & = -m\omega^{2} x=\{ p_x , H_{\alpha}\}_{\alpha},\\
&\dot y & = {p_y}/{m}=\{ y , H_{\alpha}\}_{\alpha},\quad
&\dot p_y & = -m\omega^{2} y=\{ p_y , H_{\alpha}\}_{\alpha}, 
\label{eq:ecm}
\end{array}
\end{eqnarray}
i.e., this is a continuum of Hamiltonian structures for
the two-dimensional isotropic harmonic oscillator. 
The cases $\alpha=\pi/2$ and  $\alpha=\pi$ were recently
reported in Ref. \cite{merced:gnus}. Notice that for
$\alpha \not= (2n+1)\pi/2$, with $n$ an integer, after
quantizing the system we will have a noncommutative 
space in the coordinates.\\

It can be shown that also the Hamiltonian structure given 
by the Hamiltonian
\begin{eqnarray}
H_{\beta}&=&\Big[ \frac{1}{2m}(p_{y}^{2}-p_{x}^{2}) 
+\frac{m\omega^2}{2} (y^2 - x^2)\Big]\sin\beta \nonumber\\
&+&\Big[\frac{p_x p_y}{m} + m \omega^2 x y\Big]\cos\beta,
\label{eq:Ha}
\end{eqnarray}
with Poisson brackets
\begin{eqnarray}
\begin{array}{lllll}
&\{ x , p_x \}_{\beta} & = -\sin\beta, 
&\quad\{ x , p_y \}_{\beta} & = \cos\beta,\\
&\{ y , p_x \}_{\beta} & = \cos\beta, 
&\quad\{ y , p_y \}_{\beta} & = \sin\beta, \label{eq:a}
\end{array}
\end{eqnarray}
and zero otherwise, yields the equation of motion (\ref{eq:ecm})
for each $\beta$. The case $\beta=0$ was also reported in Ref. 
\cite{merced:gnus} (see also Ref. \cite{Maria:gnus}).\\

Another Hamiltonian structure yielding the same equations
of motion (\ref{eq:ecm}) is formed by the Hamiltonian
\begin{eqnarray}
H_{\gamma}&=& -\Big[ \frac{1}{2m}(p_{y}^{2}+p_{x}^{2}) 
+\frac{m\omega^2}{2} (y^2 +x^2)\Big]\sinh\gamma\nonumber\\
&+&\Big[\frac{p_x p_y}{m} + m \omega^2 x y\Big]\cosh\gamma,
\label{eq:Haa}
\end{eqnarray}
with Poisson brackets
\begin{eqnarray}
\begin{array}{lllll}
&\{ x , p_x \}_{\gamma} & = \sinh\gamma, 
&\quad \{ x , p_y \}_{\gamma} & = \cosh\gamma,\\
&\{ y , p_x \}_{\gamma} & = \cosh\gamma, 
&\quad \{ y , p_y \}_{\gamma} & = \sinh\gamma, 
\label{eq:aa}
\end{array}
\end{eqnarray}
and zero otherwise. Once again, this structure yields 
the equations of motion (\ref{eq:ecm}) for every
$\gamma$.\\

Now, for each $\alpha$, $\beta$ and $\gamma$, the quantities
\begin{eqnarray}
D&=&\left(\frac{p_x p_y}{m}+m\omega^{2} xy\right),\\
L&=&xp_y-yp_x ,\\
H_{\pm}&=&\frac{1}{2m}(p_{y}^{2}\pm p_{x}^{2}) 
+\frac{m\omega^2}{2} (y^2 \pm x^2).
\end{eqnarray}
are conserved. It can be seen that they are, in fact, the 
symmetry generators. For each $\alpha$, the algebra of these 
generators is
\begin{eqnarray}
\{D,L\}_{\alpha}&=&2 H_{+}\sin\alpha,\label{eq:sm1}\\
\{D,H_{\pm}\}_{\alpha}&=& 2\omega H_{\mp}\cos\alpha + (1\mp 1) 
\omega^{2}L\sin\alpha,\\
\{L,H_{\pm}\}_{\alpha}&=&-(1\pm 1)D\sin\alpha,\\
\{H_{+},H_{-}\}_{\alpha}&=&2\omega D\cos\alpha,
\label{eq:sm2}
\end{eqnarray}
whereas for each $\beta$,
\begin{eqnarray}
\{D,L\}_{\beta}&=&2 H_{+}\sin\beta,\label{eq:sm11}\\
\{D,H_{\pm}\}_{\beta}&=&(1\pm 1)\omega^{2}L\sin\beta,\\
\{L,H_{\pm}\}_{\beta}&=&2 H_{\mp}\cos\beta-(1\pm 1)D\sin\beta,\\
\{H_{+},H_{-}\}_{\beta}&=&-2\omega^{2}L\cos\beta,
\label{eq:sm22}
\end{eqnarray}
and for each $\gamma$,
\begin{eqnarray}
\{D,L\}_{\gamma}&=&-2 H_{-}\sinh\gamma,\label{eq:sm111}\\
\{D,H_{\pm}\}_{\gamma}&=&(1\mp 1)\omega^{2}L\sinh\gamma,\\
\{L,H_{\pm}\}_{\gamma}&=&2 H_{\mp}\cosh\gamma-(1\mp 1)D\sinh\gamma,\\
\{H_{+},H_{-}\}_{\gamma}&=&2\omega^{2}L\cosh\gamma.
\label{eq:sm222}
\end{eqnarray}
Notice that as the Lie algebra of the constants of motion
depends on the values of $\alpha$, $\beta$ and $\gamma$;
for each of these values we will have a different symmetry
group. As an example, for $\alpha=0$ the Hamiltonian
$H_{\alpha=0}$ and its corresponding Poisson brackets are 
invariant under the rotations defined by Eq. (\ref{eq:ro1}) 
and also under the scaling transformations
\begin{equation}
x' =e^{\lambda }x, \qquad 
y' =e^{-\lambda }y,\label{eq:es}
\end{equation} 
which are particularly interesting because they mix long and 
short scales together. The rotations are generated by 
$H_{+}/\omega$ and the scaling symmetry of Eq. (\ref{eq:es}) 
by $-D/\omega$. Notice that once $\alpha\not =0$, the rotational
invariance is lost. The Hamiltonian structures labeled $\beta$
and $\gamma$ are not rotationally invariant either. However,
in the case $\beta=\gamma=0$, the structure is invariant under 
the scaling transformations from Eq. (\ref{eq:es}) which are 
generated by $L$. In the case $\beta=\pi/2$, the Hamiltonian 
structure is invariant under the transformations
\begin{eqnarray}
\begin{array}{ll}
& x' = x\cosh\theta+y\sinh\theta,\\
& y' = x\sinh\theta+y\cosh\theta.\label{eq:rro2}
\end{array}
\end{eqnarray} 
That is, this structure posses the Lorentz symmetry. For this, 
the generator of the symmetry is $L$. Notice here that the 
invariant associated to these transformations is $s^{2}=x^{2}-y^{2}$. 
On the other hand, it can be shown that the generators of
rotations for $\beta=0$ and $\beta=\pi$ are $L_{1}=xp_x-yp_y$
and $L_{2}=xp_y+yp_x$ respectively. Both of these are not
conserved quantities; which confirms the fact that both systems
have no rotational symmetry. The alternative Hamiltonian structures
here presented have particular symmetries and, in this sense,
they are all different.\\

In general, when the equations of motion of a physical system 
can be obtained from alternative Hamiltonian structures, it 
is difficult to argue why one choses the standard structure
over the others \cite{ok:gnus,lemos:gnus,u:gnus}. For the 
two-dimensional isotropic harmonic oscillator and the 
alternative structures here presented, this is not so
complicated. As explained above, the case $\alpha=0$ can be 
clearly discarded because both $H_{\alpha=0}$ and its 
corresponding Poisson brackets lose meaning in the limit
$\omega\to 0$. The other cases for $\alpha$, $\beta$ and
$\gamma$ can, in principle, be discarded because they have no
rotational symmetry. Thus, the best mathematical representation 
for this physical system is provided by the usual Hamiltonian 
structure. This, however, does not mean one should fully neglect 
the others as they may give a better representation of other 
systems and, in addition, may have interesting properties. For 
instance, in the Hamiltonian structure labeled $\alpha$ above, if
$\cos\alpha\not= 0$ we will have noncommutative spaces. That
is, we will have the commutation rules
\begin{equation}
[\hat x,\hat y]=i\hbar\Theta, \quad 
\Theta=\frac{\cos\alpha}{m\omega}.
\end{equation}
Moreover, if $\cos\alpha>0$, it can be shown that with these
commutation rules, the quantity
$\hat A=\pi \hat r^{2}=\pi(\hat x^{2}+\hat y^{2})$,
has the spectrum \cite{y:gnus,yy:gnus},
\begin{eqnarray}
A_{N}=2\pi\hbar\Theta(N+1/2),\quad  N=0,1,2,\dots \label{eq:Area}
\end{eqnarray}
i.e. the area is quantized. Some authors \cite{bk:gnus} believe 
that the area of the horizon of events of a black hole has a 
spectrum of the form of Eq. (\ref{eq:Area}). The Hamiltonian
structures invariant under the scaling transformations from
Eq. (\ref{eq:es}) are also interesting because they mix different
scales together. It is also notable that the structure with
$\beta=\pi/2$ posses the Lorentz symmetry despite it is a
non relativistic model.

\begin{acknowledgments}
The authors would like to thank ICN-UNAM for its kind hospitality
and AZ also for financial support.
\end{acknowledgments}


\begin{thebibliography}{9}
%
\bibitem{Maria:gnus}
R.~M.~Santilli, {\it Foundations of Theoretical Mechanics. The Inverse 
Problem in Newtonian Mechanics} (Springer-Verlag, New York, 1978), Part I.

\bibitem{wigner:gnus}
E.~P.~Wigner, {\it Do the Equations of Motion Determine the 
Quantum Mechanical Commutation Relations?} Phys. Rev. {\bf 77}, 711 (1950).

\bibitem{ok:gnus}
S.~Okubo, {\it Does the equation of motion determine commutation
relations?} Phys. Rev. D {\bf 22}, 919 (1980).

\bibitem{shepley:gnus}
M.~Henneaux and L.~C.~Shepley, {\it Lagrangians for spherically symmetric
potentials,} J. Math. Phys. {\bf 23}, 2101 (1982).

\bibitem{lemos:gnus}
N.~A.~Lemos, {\it Physical consequences of the choice of the Lagrangian,}
Phys. Rev. D {\bf 24}, 1036 (1981).

\bibitem{u:gnus}
S.~A.~Hojman and L.~F.~Urrutia,
{\it Comments on "Physical consequences of the choice of the Lagrangian"}
Phys. Rev. D {\bf 26}, 527 (1982).

\bibitem{merced:gnus}
M.~Montesinos and G.~F.~Torres del Castillo, {\it Symplectic quantization, 
inequivalent quantum theories, and Heisenberg's principle of uncertainty,}
Phys. Rev. A {\bf 70}, 032104 (2004), quant-ph/0407051.

\bibitem{y:gnus}
J.~M.~Romero, J.~A.~Santiago and J.~D.~Vergara,
{\it A note about the quantum of area in a non-commutative space,}
Phys. Rev. D {\bf 68}, 067503 (2003), hep-th/0305080.

\bibitem{yy:gnus}
S.~Sivasubramanian, G.~Srivastava, A.~Vitiello and Y.~N.~Widom, 
{\it Quantum dissipation induced noncommutative geometry,} 
Phys. Lett. A {\bf 311}, 97 (2003), quant-ph/0301005.

\bibitem{bk:gnus}
J.~D.~Bekenstein, {\it Quantum Black Holes as Atoms,} gr-qc/9710076;
A.~Alekseev, A.~P.~Polychronakos, and M.~Smedb\"ack, 
{\it On area and entropy of a black hole,} Phys. Lett. B {\bf 574}, 
296 (2003), hep-th/0004036.


%
\end{thebibliography}
\end{document}